\documentclass[authoryear,preprint,review,12pt]{elsarticle}

\usepackage{amssymb}
\usepackage{amsthm}
\usepackage{amsmath}
\usepackage{adjustbox}
\usepackage{setspace}
\usepackage{longtable}
\usepackage{booktabs}
\usepackage{xcolor}
\usepackage{lscape}
\usepackage{tabularx}
\usepackage{booktabs}
\usepackage{array}
\usepackage{tikz}
\usepackage{tikz}
\usepackage{graphicx}
\usepackage{amsmath, amssymb} 
\usetikzlibrary{shapes.geometric, arrows.meta, positioning, fit, calc, shadows}
\usepackage{float}            
\usepackage{rotating}
\usepackage{lscape}
\usepackage{booktabs}
\usepackage{soul}
\usepackage{textcomp}
\usepackage{tabu}
\usepackage{tocloft}
\usepackage{blindtext}
\usepackage{mathtools}
\usepackage{blkarray}
\usepackage{subcaption}
\usepackage{rotating}
\usepackage{lscape}
\usepackage{array}
\usepackage{booktabs}
\usepackage{soul}
\usepackage{textcomp}
\usepackage{amsmath}
\usepackage{amssymb}
\usepackage{verbatim} 
\usepackage{longtable}
\usepackage{adjustbox}
\usepackage{longtable}
\usepackage{blkarray, bigstrut}
\usepackage{breqn}
\usepackage{xparse}
\usepackage{hyperref}
\hypersetup{
			colorlinks=true,
			linkcolor=blue,
            linktoc=page,
			anchorcolor=black,
			citecolor=blue,
			urlcolor=blue}

\usepackage{leftidx}

\usepackage{lineno}
\usepackage{doi}

\begin{document}

\begin{frontmatter}



\title{Community detection in small-sample ordinal regimes: A benchmarking framework for Delphi data}


\author[inst1]{Yuri Calleo}
\affiliation[inst1]{Department of Economics, Statistics and Business, Universitas Mercatorum, Rome}

\author[inst2]{Simone Di Zio}
\affiliation[inst2]{Department of Socio-Economic, Managerial, and Statistical Studies, University "G. d'Annunzio", Chieti-Pescara}
            
\author[inst1]{Fabrizio Maturo}

\begin{abstract}
 The statistical modeling of consensus in Delphi data faces a critical bottleneck: the high dimensionality of questionnaire items relative to the limited sample size of expert panels. This rank deficiency leads traditional latent variable models, such as Principal Component Analysis, to be structurally unstable and prone to overfitting. Addressing this methodological gap, this study proposes a transition from variable-centric covariance models to network-centric connectivity models. By mapping item correlations onto a weighted graph topology, we present a simulation-based benchmark that utilizes community detection algorithms to identify latent thematic structures, effectively addressing the spectral instability and rank deficiency typical of high-dimensional, low-sample-size regimes. The research systematically evaluates the robustness of topological approaches based on structural density, information flow, and spectral partitioning against synthetic datasets designed to replicate the pathological conditions of consensus data, including ordinal scales and systemic noise. The central methodological contribution lies in demonstrating that collinearity among expert judgments - traditionally treated as statistical redundancy to be regularized - can be effectively reinterpreted as a topological signal of cohesion. This framework provides researchers with a structured and automated procedure for dimensionality reduction, ensuring structural stability and psychometric consistency even in small-sample regimes where standard factor analysis breaks down.
\end{abstract}




\end{frontmatter}

\section{Introduction}
\label{Introduction}

The statistical modeling of high-dimensional data presents a formidable challenge in computational statistics, particularly in regimes characterized by a low subject-to-item ratio ($N:P$). Ideally, multivariate techniques such as Exploratory Factor Analysis (EFA) or Principal Component Analysis (PCA) require a sample size significantly larger than the number of variables - typically a ratio of at least 5:1 or 10:1 - to yield replicable results \citep{Watkins2018}. When this ratio drops towards 1:1 or 2:1 ("Small $N$"), the estimation of the covariance structure enters a pathological domain. In this regime, even if the sample covariance matrix is mathematically invertible, it becomes statistically unstable: the sample eigenvalues diverge systematically from the true population eigenvalues, treating sampling noise as signal. This invalidates the asymptotic properties required for classical dimensionality reduction, leading to solutions that are often non-replicable or mathematical artifacts. A paradigmatic example of this data regime is the Delphi method, a structured communication technique that relies on carefully selected panels to aggregate expert judgment into collective consensus \citep{Rowe1999}. While the procedural dynamics of the Delphi are well-documented, the statistical treatment of items faces severe constraints. To operationalize consensus, experts typically cast judgments on $P \in [20, 100]$ items, with panel sizes rarely exceeding $N=40-50$ due to logistical barriers \citep{Baker2006}. In this context, where the $N:P$ ratio is critically low, standard Latent Variable Models (LVMs) may lack sufficient statistical power. The resulting factor structures are prone to massive overfitting \citep{Zhang2022DelphiAHP}. While the statistical literature offers advanced solutions to mitigate these issues - such as shrinkage covariance estimators, Sparse PCA, and Ordinal Factor Models designed to handle discrete distributions and rank-deficient matrices - these techniques inherently retain a continuous latent-variable ontology. Consequently, they still require subjective thresholding to translate continuous factor loadings into the discrete, bounded thematic clusters often required for practical Delphi applications. Consequently, standard analytical protocols revert to univariate reductionism, decomposing the multivariate space into $P$ independent problems and assessing consensus via scalar metrics such as the Median ($Mdn$) or Interquartile Range ($IQR$) \citep{ZarthaSossa2019, Author2021}. However, this reductionist approach - although useful for preliminary analysis - relies on the strong, often violated assumption of local independence, treating each item as an orthogonal entity \citep{Author2024}. By ignoring the joint probability distribution of responses, researchers discard critical information regarding the semantic structure of the consensus - specifically, the conditional probability that an expert endorses item $i$ given their endorsement of item $j$. As noted by \cite{Author2025}, this creates a methodological paradox: the Delphi method aims to synthesize a complex, holistic view of a problem, yet its statistical tools fragment this view into isolated, uncorrelated components. The limitations in applying Latent Variable Models (LVMs) like Principal Component Analysis (PCA) due to the rank deficiency of the sample covariance matrix forces a loss of theoretical coherence, potentially retaining redundant or collinear items that distort the final consensus. 

To resolve the impasse posed by rank-deficient ordinal matrices, we propose a topological shift in the analytical approach: moving from variable-centric covariance models to network-centric connectivity models. By mapping the regularized correlation matrix to a weighted graph topology, we can provide a robust estimation even in ill-conditioned scenarios, avoiding spectral instability, and adopt techniques from Network Science - specifically community detection - to identify latent modular structures \citep{LancichinettiFortunato2009}. In this framework, collinearity is reinterpreted not as a statistical defect (singularity), but as a structural signal of connectivity, allowing for the identification of coherent "belief systems" even in sparse data regimes \citep{Dormann2013}.

In proposing this topological shift, it is essential to precisely define the inferential goal of our framework. Rather than acting as a generic dimensionality reduction tool, this network-based approach is specifically intended to solve the problem of feature clustering for the identification of latent thematic structures under severe sample constraints. In the context of Delphi research, extracting these communities is not an end in itself, but serves a dual operational purpose. First, for consensus measurement, the topological clusters identify highly reliable sub-scales (as measured by internal consistency) that can be aggregated into composite indicators. Second, for scenario formation - a primary objective in foresight studies - the network partitions ensure that interdependent items are grouped into coherent, statistically validated narrative blocks, preventing the cognitive overload associated with interpreting dozens of isolated expert judgments. Therefore, the resulting communities should be interpreted as robust, emergent "belief systems" shared by the expert panel.

While network analysis is gaining traction in scientific literature \citep{Borsboom2021}, a critical gap remains: the choice of the clustering algorithm for Delphi data is often methodologically underspecified, as algorithms are frequently imported from other domains (e.g., social network analysis or biology) without a formal evaluation of their suitability for rank-deficient, ordinal matrices \citep{Tapio2003PolicyDelphi, Author2025}. 

Furthermore, in the existing Delphi literature, network analysis is employed almost exclusively as a descriptive tool to visualize expert interactions or consensus patterns (see for an extensive review \citep{ZarthaSossa2019}. It has not yet been formally framed as a structural substitute for latent variable models (such as EFA) in small-sample regimes. This lack of a specific selection criterion is problematic because different algorithms optimize different objective functions - from static density (Modularity) to dynamic information flow (Map Equation) - and it is currently unknown which topology best represents the latent structure of expert consensus under the constraints of small - sample ordinal data (such as the Delphi method) \citep{Saffie2016FuzzyDelphi} Consequently, the lack of a benchmarking standard creates an epistemological ambiguity: without determining whether expert consensus manifests topologically as cohesive density or dynamic information flow, researchers cannot distinguish between genuine latent belief systems and mere algorithmic artifacts \citep{Schaub2017CommunityDetection}.
This paper addresses this gap by conducting a simulation-based benchmark of community detection algorithms for small-sample ordinal Delphi data. Specifically, we aim to:
\begin{enumerate}
    \item Formalize the translation of rank-deficient ordinal matrices into weighted, undirected networks using regularized association measures.
    \item Benchmark the performance of three distinct classes of community detection algorithms - Modularity maximization (Louvain, Leiden), Flow-based dynamics (Infomap), and Spectral partitioning (Ncut) - in the specific context of expert consensus data.
    \item Validate the topological approach through a rigorous simulation framework, quantifying both ground-truth recovery (ARI) and psychometric internal consistency ($\alpha$) under conditions of unbalanced latent structures (with a sensitivity analysis protocol).
\end{enumerate}

The primary contribution of this work is to provide a practical alternative to latent variable models for high-dimensional, small-sample consensus data. This approach is introduced to resolve the specific issue of clustering expert judgments for the generation of future scenarios.
The remainder of the paper is organized as follows: Section \ref{sec:technicalframework} outlines the technical framework, contrasting the linear algebra of latent factors with the topology of network models. Section \ref{sec:methodology} details the methodology and the simulation protocol. Section \ref{sec:simulations} presents the results, benchmarking the algorithms' ability to recover latent themes from noisy, discretized data. Finally, Section \ref{sec:conclusion} discusses the implications for quantitative consensus analysis.

\section{Technical Framework}
\label{sec:technicalframework}

\subsection{The dimensionality problem: latent variables vs. network topology}
Traditionally, the dimensionality reduction of Delphi questionnaires may be approached through the lens of the Common Cause Model, typically implemented via Exploratory Factor Analysis (EFA) or Principal Component Analysis (PCA) \citep{Fleming1986}. Let $\mathbf{X} \in \mathbb{R}^{N \times P}$ be the matrix of observed ratings for $N$ experts and $P$ items. The latent variable model assumes that the observed variance is generated by a low-dimensional vector of unobserved factors $\boldsymbol{\xi} \in \mathbb{R}^F$ ($F \ll P$):

\begin{equation}
    \mathbf{x} = \boldsymbol{\Lambda} \boldsymbol{\xi} + \boldsymbol{\epsilon}
\end{equation}

\noindent where $\boldsymbol{\Lambda}$ is the factor loading matrix and $\boldsymbol{\epsilon}$ is the measurement error (unique variance). A fundamental axiom of this model is local independence: given the latent factors, the observed variables must be statistically independent (i.e., the off-diagonal elements of the residual covariance matrix must be zero) \citep{BerenschotGrift2019}. Under this common-cause framework, correlations among observed items are not interpreted as direct interactions between items, but as spurious associations induced by shared latent factors, such that, conditional on the latent structure, any residual dependence among items is assumed to vanish. Indeed, this framework faces a critical stability issue in the context of Delphi datasets, since they are characteristically defined by a "small sample" regime, where the number of experts ($N$) sometimes is comparable to the number of items ($P$). In this regime, the statistical properties of the sample covariance matrix $\mathbf{S} = \frac{1}{N-1}\mathbf{X}^T\mathbf{X}$ are heavily dictated by the concentration ratio $P/N$. It is crucial to distinguish between two related pathologies depending on this ratio. When $N \le P$, the covariance matrix is strictly rank-deficient and mathematically singular, meaning its inverse (the precision matrix $\mathbf{\Theta} = \mathbf{S}^{-1}$) does not exist. This strict singularity directly precludes the use of likelihood-based Latent Variable Models (such as Maximum Likelihood Exploratory Factor Analysis), which rely on matrix inversion to evaluate model fit. On the other hand, when $N$ is only moderately larger than $P$, the matrix may be full rank but becomes highly ill-conditioned. While PCA circumvents direct inversion by relying on eigen decomposition or Singular Value Decomposition (SVD), it is not immune to the small-sample regime. As the $P/N$ ratio grows, the sample covariance matrix suffers from severe spectral distortion. The sample eigenvalues systematically diverge from the true population eigenvalues—treating random sampling fluctuations as structural variance. This distortion invalidates standard asymptotic properties, causing dimensionality retention heuristics (such as Kaiser's rule or Parallel Analysis) to fail, ultimately leading to mathematical artifacts and massive overfitting \citep{Bokrantz2017}.

It must be acknowledged that the statistical literature provides advanced extensions of variance-based models designed specifically to address the $N < P$ regime and the discrete nature of Likert-type scales. Techniques such as Shrinkage PCA \citep{chen2010denoising} or Ridge-regularized PCA \citep{warton2008penalized} stabilize the condition number of the sample covariance matrix by applying penalty terms, preventing mathematical singularity. Furthermore, Ordinal Factor Models \citep{Grilli2007} explicitly account for the discrete thresholds of the response scale, avoiding the distortion of Pearson correlations. However, while these regularized and ordinal alternatives successfully resolve the algebraic and distributional challenges of LVMs, they do not fully align with the applied goal of feature clustering. Unless strict sparsity constraints are imposed (e.g., Sparse PCA), these models yield continuous loadings across multiple latent dimensions.

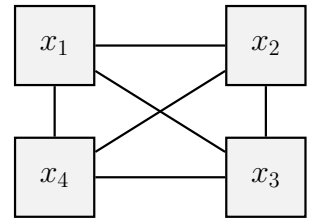
\begin{figure}[h]
\centering
\resizebox{\textwidth}{!}{
\begin{tikzpicture}[
    latent/.style={circle, draw=black, very thick, minimum size=1.8cm, align=center, font=\Large},
    obs/.style={rectangle, draw=black, fill=gray!10, very thick, minimum size=1.5cm, align=center, font=\Large},
    arrow/.style={->, >=stealth, line width=1.5pt},
    label_title/.style={font=\bfseries\Large, align=center},
    label_desc/.style={text width=6cm, align=center, font=\normalsize, text=black!80}
]


\node[latent] (xi) at (0, 3.0) {$\xi$}; 
\node[obs] (x1) at (-3.0, 0) {$x_1$};
\node[obs] (x2) at (-1.0, 0) {$x_2$};
\node[obs] (x3) at ( 1.0, 0) {$x_3$};
\node[obs] (x4) at ( 3.0, 0) {$x_4$};

\draw[arrow] (xi) -- (x1);
\draw[arrow] (xi) -- (x2);
\draw[arrow] (xi) -- (x3);
\draw[arrow] (xi) -- (x4);

\node[label_title] at (0, -2.0) {(a) Common Cause Model};
\node[label_desc] at (0, -3.5) {Variables are locally independent given the latent factor ($\xi$).};

\begin{scope}[xshift=11cm]

    \node[obs] (n1) at (-2.0, 2.0) {$x_1$};
    \node[obs] (n2) at ( 2.0, 2.0) {$x_2$};
    \node[obs] (n3) at ( 2.0, -0.5) {$x_3$};
    \node[obs] (n4) at (-2.0, -0.5) {$x_4$};

    \draw[very thick] (n1) -- (n2);
    \draw[very thick] (n2) -- (n3);
    \draw[very thick] (n3) -- (n4);
    \draw[very thick] (n4) -- (n1);
    \draw[very thick] (n1) -- (n3); 
    \draw[very thick] (n2) -- (n4);
    \node[label_title] at (0, -2.0) {(b) Pairwise Network (PMRF)};
    \node[label_desc] at (0, -3.5) {Emergent constructs arise from direct mutual interactions.};

\end{scope}

\end{tikzpicture}} 
\caption{Conceptual comparison between the traditional Latent Variable Model (a) and the Network Perspective (b). In the latent model, correlations are spurious, caused by $\xi$. In the network model, correlations are constitutive of the construct itself.}
\label{fig:model_comparison}
\end{figure}

To overcome the limitations of rank deficiency and spectral instability inherent in the $N \ll P$ regime, we consider a Network Perspective \citep{Borsboom2021} (see Fig. \ref{fig:model_comparison}). This approach rejects the local independence assumption, and instead of viewing items as passive indicators of a latent cause, it models them as active entities. In this ontology, constructs are not antecedent causes but emergent properties of a mutualism model: items cluster together because they directly reinforce each other (e.g., in a decision-making context, assessing a risk as "high" may causally increase the probability of rating a mitigation strategy as "necessary").

Formally, we can define the Delphi consensus structure as a graph $G = (V, E, W)$, where:
\begin{itemize}
    \item $V = \{1, \dots, P\}$ is the set of nodes (items).
    \item $E \subseteq V \times V$ is the set of edges representing conditional dependence.
    \item $W: E \rightarrow \mathbb{R}$ is a mapping assigning weights based on the strength of association (e.g., regularized partial correlations or rank correlations).
\end{itemize}
Edges in the proposed network representation are undirected, reflecting symmetric statistical associations between items rather than directional or causal relationships. Furthermore, we do not assume a fully connected topology; rather, we adopt a sparse network structure where the existence of an edge is conditional on statistical evidence. Specifically, weights are set to zero ($A_{ij}=0$) unless the association satisfies a rigorous significance criterion (e.g., FDR correction), thereby filtering out sampling noise. This approach ensures that the resulting topology captures only the robust backbone of the expert consensus, consistent with the ordinal and small-sample nature of Delphi data.
The analytical objective shifts from estimating factor loadings $\boldsymbol{\Lambda}$ to identifying the optimal partition $\mathcal{C} = \{C_1, \dots, C_K\}$ of the vertex set $V$, such that the topology of the graph is maximized according to specific community detection criteria. Among the several possible analytical targets in a network representation (e.g., centrality, connectivity patterns, or node-level roles), we deliberately focus on community detection, as the identification of coherent item clusters directly aligns with the goal of extracting thematic structures and consensus dimensions in Delphi questionnaires. From a taxonomic perspective, transitioning to a network model implies a structural shift from dimensionality reduction (or feature transformation) to feature clustering. While standard latent variable models like PCA create new synthetic variables via linear combinations, community detection partitions the existing feature space, identifying cohesive groups of correlated items without altering their original ordinal nature. We maintain a benchmarking comparison with PCA because it represents the deeply entrenched status quo in applied Delphi research, where factor loadings are routinely misused to manually cluster items. However, within the proper feature clustering family, the proposed network framework offers significant structural advantages over traditional methods like Hierarchical Agglomerative Clustering (HAC) \citep{murtagh2014ward}. While HAC relies on the subjective visual inspection of a dendrogram to establish the final item partitions, topological algorithms optimize global structural objective functions (e.g., Modularity or the Spectral Eigengap), providing an automated, mathematically grounded criterion for determining the optimal number of consensus themes. Moreover, network-based community detection offers additional advantages over traditional clustering methods: it is more robust to collinearity and noise, optimizes global topological criteria, handles unbalanced cluster sizes, and operates on statistically filtered sparse networks, resulting in more stable and interpretable partitions. The interpretability of the resulting groups is crucial in the passage from the Delphi outputs to decisions or to building future scenarios.

\subsection{Algorithmic classes in community detection}
The choice of the community detection algorithm is not merely computational but imposes specific topological assumptions on the definition of "consensus". In this study, we benchmark three distinct theoretical classes: (a) Modularity maximization, (b) Flow-based methods, and (c) Spectral methods.

(a) The first class of algorithms views consensus as "structural density". In this perspective, consensus is defined as groups of items that are more densely interconnected with each other than with the rest of the network. They attempt to maximize the modularity function $Q$, which measures the concentration of edges within communities relative to a random null model (typically the configuration model preserving degree distribution) \citep{Traag2019}:

\begin{equation}
   Q = \frac{1}{2m} \sum_{ij} \left( A_{ij} - \frac{d_i d_j}{2m} \right) \delta(c_i, c_j)
\end{equation}

\noindent where $A_{ij}$ is the adjacency weight, $d_i$ is the node degree, $m$ is the total weight, and $\delta$ is the Kronecker delta. This objective function serves as a global quality metric that can be optimized through different heuristic algorithms.
We employ the Louvain algorithm, a greedy optimization heuristic and applied for the first time to Delphi data in \cite{Author2025}. However, Louvain is known to produce disconnected communities and suffer from the "resolution limit" (merging small, distinct clusters into larger ones). To address these topological defects, we also include the Leiden algorithm, which refines the local move phase to guarantee the connectivity of communities and improve the resolution of sub-structures \citep{Traag2019}.

(b) Unlike modularity, which relies on static density, flow-based methods (i.e., Infomap) model the consensus as a dynamic system of information flow. Here, information flow is not intended as a real temporal or causal process, but as a conceptual device to identify regions of the network where transitions tend to remain confined. It minimizes the map Equation, which represents the Shannon entropy of a random walker's trajectory on the network \citep{AlzahraniHoradam2015}. The objective is to minimize the description length $L(M)$ of the random walk:

\begin{equation}
    L(M) = q_\curvearrowright H(\mathcal{Q}) + \sum_{c=1}^{K} p_{\circlearrowright}^c H(\mathcal{P}^c)
\end{equation}

The first term, $q_\curvearrowright H(\mathcal{Q})$, measures the information cost of moving between modules, while the second term captures the cost of moving within modules. In a Delphi context, Infomap can be particularly sensitive to "cognitive pathways": if experts conceptually associate \textit{Item A} with \textit{Item B}, a random walker could tend to stay trapped in that cluster. This method could be theoretically superior for detecting nested sub-themes that might be structurally embedded within larger dense regions.

(c) Finally, spectral methods relax the discrete clustering problem into a continuous optimization problem using the eigenvalues (spectrum) of the Graph Laplacian matrix. In this perspective, consensus is defined as a partition that separates the network into the most stable components, minimizing the impact of small perturbations or noise. The graph Laplacian provides a spectral representation of the network that encodes how naturally the graph can be partitioned. We utilize the Normalized Cut (Ncut) objective, solved via the eigendecomposition of the normalized Laplacian \citep{Li2018}. 

\begin{equation}
    \mathbf{L}_{sym} = \mathbf{I} - \mathbf{D}^{-1/2}\mathbf{A}\mathbf{D}^{-1/2}
\end{equation}

\noindent where $\mathbf{I}$ is the identity matrix, $\mathbf{A}$ represents the weighted adjacency matrix, and $\mathbf{D}$ is the diagonal degree matrix, with elements defined as the sum of incident weights ($D_{ii} = \sum_j A_{ij}$). A key advantage of spectral methods is that the eigenvalues of the Laplacian contain information about the number of meaningful clusters present in the network.

However, a critical challenge in spectral clustering is determining the number of communities a priori. To resolve this in an unsupervised setting, we rely on the Matrix Perturbation Theory and the Eigengap Heuristic (based on the Davis-Kahan theorem). This heuristic posits that the optimal $K$ corresponds to the maximum difference between consecutive eigenvalues ($\delta_r = |\lambda_{r+1} - \lambda_r|$), representing the most stable separation of the graph's eigenspace before introducing noise. Intuitively, the eigengap identifies the point beyond which further partitions would reflect noise rather than genuine structure (the optimal number of clusters $K^*$ is determined by maximizing this spectral gap).

\section{Methodology}
\label{sec:methodology}

\subsection{Data formalization and network construction}
\label{subsec:data_formalization}

Let $\mathcal{X}^{(r)}$ be the $N \times P$ response matrix for round $r$ of a Delphi study, where $x_{ij} \in \mathbb{S}$ denotes the rating of the $i$-th expert ($i=1, \dots, N$) on the $j$-th item ($j=1, \dots, P$) drawn from an ordinal scale $\mathbb{S}$ (e.g., $\{1, \dots, 10\}$). In this study, we focus on the analysis of the final round; however, the method is flexible for applications in different or multiple rounds. The column vector $\mathbf{x}_{\cdot j}$ represents the vector of ratings for item $j$. 

We define the weighted adjacency matrix $\mathbf{W} \in \mathbb{R}^{P \times P}$ where each element $w_{ij}$ quantifies the statistical association between items $j$ and $l$. Given the ordinal and often non-normal nature of Delphi data, we utilize the Spearman rank-order correlation ($\rho$) \citep{Zar1972}. For Spearman's $\rho$, let $R_{ij}$ be the rank of $x_{ij}$ across experts within item $j$.  To ensure consistency, given the discrete nature of the Likert scale, ties (identical ratings) are resolved using the average rank method \citep{Conover1999}. Then, the correlation between item $j$ and item $l$ is computed as:

\begin{equation}
    w_{jl} = \rho(\mathbf{x}_{\cdot j}, \mathbf{x}_{\cdot l}) = \frac{\sum_i (R_{ij} - \bar{R}_j)(R_{il} - \bar{R}_l)}{\sqrt{\sum_i (R_{ij} - \bar{R}_j)^2 \sum_i (R_{il} - \bar{R}_l)^2}}
\end{equation}

\noindent where $\overline{R}$ represents the average rank within a column. This equation calculates the normalized covariance of the ranks, effectively treating the rank vectors $\mathbf{x}_{\cdot j}$ and $\mathbf{x}_{\cdot l}$ as input for the Pearson correlation coefficient.

Given the correlation metric is symmetric ($w_{jl} = w_{lj}$), the resulting network is inherently undirected, consistent with the framework defined in Section \ref{sec:technicalframework}. Furthermore, as standard modularity maximization typically requires non-negative weights, we transform $\mathbf{W}$. While negative correlations in Delphi data (indicating trade-offs or disagreement) are informative, the primary goal is often identifying coherent sub-themes of positive consensus; therefore, negative associations are excluded, as they do not support the aggregative notion of community underlying standard modularity-based clustering. We strictly separate positive coherence by defining the non-negative adjacency matrix ${\mathbf{A^+}}$:

\begin{equation}
    A^+_{jl} = \begin{cases} 
    w_{jl} & \text{if } w_{jl} > \tau \\
    0 & \text{otherwise}
    \end{cases}
\end{equation}

\noindent where $\tau \ge 0$ is a threshold parameter. To avoid arbitrary thresholding, we employ a statistical significance filter, keeping edges where $p < .05$ with False Discovery Rate (FDR) correction \citep{Benjamini1995}. Finally, to ensure topological consistency, isolated nodes (degree $d=0$) resulting from the filtering process are pruned from the network before clustering.

\subsection{Community detection protocols}
\label{subsec:community_detection}

To benchmark the efficacy of the topological approach, we apply the three distinct classes of algorithms defined in Section \ref{sec:technicalframework} to the weighted adjacency matrix ${\mathbf{A^+}}$. First, for Modularity-based optimization, we employ both the classical Louvain algorithm and the refined Leiden algorithm. Given that the optimization of modularity $Q$ is NP-hard, these algorithms rely on a heuristic search that iteratively moves nodes to neighboring communities to maximize the local gain. Specifically, for a node $i$ moving into community $C$, the change in modularity $\Delta Q$ is calculated as:

\begin{equation}
   \Delta Q = \left[ \frac{\Sigma_{in} + 2d_{i,in}}{2m} - \left( \frac{\Sigma_{tot} + d_i}{2m} \right)^2 \right] - \left[ \frac{\Sigma_{in}}{2m} - \left( \frac{\Sigma_{tot}}{2m} \right)^2 - \left( \frac{d_i}{2m} \right)^2 \right]
\end{equation}

\noindent where $\Sigma_{in}$ is the sum of weights inside $C$, $\Sigma_{tot}$ is the sum of weights of edges incident to nodes in $C$, $d_i$ is the degree of node $i$, and $d_{i,in}$ is the sum of weights from $i$ to nodes in $C$. 

Second, for the Flow-based approach, we deploy the Infomap algorithm. We model the network as a static, undirected, and weighted graph where the flow of the random walker is guided by the edge weights derived from the Delphi correlations. The transition probability $P_{ij}$ for a walker to move from item $i$ to item $j$ is proportional to the edge weight:

\begin{equation}
 P_{ij} = \frac{A^+_{ij}}{\sum_{l} A^+_{il}} = \frac{A^+_{ij}}{d_i}
\end{equation}

\noindent where \(A^+_{ij}\) denotes the positive correlation between items \(i\) and \(j\), and 
\(d_i = \sum_l A^+_{il}\) represents the total strength of all edges incident to node \(i\). 
The algorithm is configured to minimize the map equation using a two-level partition scheme, setting, with a number of outer-loop trials set to 20.

Finally, for Spectral clustering, we solve the relaxed continuous optimization problem to minimize the Normalized Cut ($Ncut$). We first compute the symmetric normalized Laplacian matrix $L_{sym}$:

\begin{equation}
   \mathbf{L}_{sym} = \mathbf{I} - \mathbf{D}^{-1/2} \mathbf{A}^+ \mathbf{D}^{-1/2}
\end{equation}

\noindent where $\mathbf{D}$ is the diagonal degree matrix with $D_{ii} = \sum_j A^+_{ij}$. 

As stated in Section \ref{sec:technicalframework}, a critical constraint of spectral methods is the requirement to pre-specify the number of clusters $K$. To address this in an unsupervised Delphi context, we employ the eigengap heuristic. We compute the eigenvalues $0 = \lambda_1 \leq \lambda_2 \leq \dots \leq \lambda_P$ of $L_{sym}$. The optimal number of communities $K^*$ is determined by maximizing the spectral gap $\delta$:

\begin{equation}
  K^* = \underset{r}{\arg\max} \, \delta_r = \underset{r}{\arg\max} \, |\lambda_{r+1} - \lambda_r|
\end{equation}

\noindent This index indicates the most stable separation of the graph's spectrum. The resulting first \(K^*\) eigenvectors are then clustered using K-means to obtain the discrete partition.

\section{Simulation study}
\label{sec:simulations}
\subsection{Simulation framework and validation}
\label{subsec:simulation}

To validate the methodology under the structural constraints typical of Delphi studies (Small $N$, Large $P$), we designed a simulation framework. The objective is to benchmark the algorithms' ability to recover a known latent structure ("Ground Truth") from noisy, ordinal data. The data generation process follows a three-stage protocol: 

(1) \textit{Latent structure definition}: we model the expert opinions as a latent random vector $\mathbf{z} \in \mathbb{R}^P$ following a multivariate normal distribution $\mathcal{N}(\boldsymbol{\mu}, \mathbf{\Sigma})$. We simulate a panel of $N=20$ experts rating $P=60$ items. The covariance matrix $\mathbf{\Sigma}$ is constructed with a block-diagonal structure to represent $K_{true}=3$ thematic clusters. Crucially, to replicate the heterogeneity of real-world consensus, we impose unbalanced cluster sizes ($P_1=10$, $P_2=20$, $P_3=30$ items). This design is specifically intended to stress-test the algorithms against the "resolution limit" phenomenon, which often causes modularity-based methods to merge small communities into larger ones. The choice of $K_{true}=3$ is therefore not meant to reflect a universal number of thematic domains in Delphi studies, but to provide a minimal yet non-trivial benchmark for evaluating algorithmic sensitivity under controlled conditions.

(2) \textit{Signal-to-noise specification}: the matrix $\mathbf{\Sigma}$ is parameterized to simulate a realistic, imperfect consensus. Specifically:
\begin{itemize}
    \item Intra-cluster correlation (Signal): items belonging to the same cluster share a correlation of $\rho_{in} = 0.50$. This represents a moderate-to-strong expert consensus.
    \item Inter-cluster correlation (Noise): items from different clusters share a baseline correlation of $\rho_{out} = 0.10$. This introduces "crosstalk" or systemic noise, mimicking the cognitive overlap often found in complex policy questionnaires.
\end{itemize}

(3) \textit{Ordinal discretizations}: since Delphi outputs cannot be considered continuous (even if expressed with numerical evaluations), we project the latent continuous values onto a discrete ordinal space $\mathbb{S} = \{1, \dots, 10\}$. The observed rating $x_{ij}$ for expert $i$ on item $j$ is obtained by applying a binning function $\phi$ to the latent value $z_{ij}$:
\begin{equation}
    x_{ij} = \phi(z_{ij}) = \sum_{b=1}^{10} b \cdot \mathbb{I}(c_{b-1} < z_{ij} \le c_b)
\end{equation}
\noindent where $c_b$ are the thresholds defining the 10 intervals of the Likert scale. This discretization step introduces information loss and quantization noise, challenging the robustness of the correlation estimation.

To empirically demonstrate the theoretical limitations of latent variable models in small-sample regimes, we implemented a direct benchmarking protocol contrasting the proposed topological framework against standard PCA \citep{ZwickVelicer1986}. The suitability of the data for factor analysis is first assessed via the Kaiser-Meyer-Olkin (KMO) measure of sampling adequacy and Bartlett's test of sphericity \citep{Tobias1969}. Subsequently, we evaluate the dimensionality recovery using a dual approach for PCA: the standard Kaiser's criterion (eigenvalues $>1$) and, to rigorously control for chance capitalization, Horn's Parallel Analysis based on Monte Carlo simulation \citep{Timmerman2011}. These are contrasted against the modularity optimization and spectral gap heuristics used for network methods. Finally, classification accuracy is compared by forcing the PCA solution to the true number of clusters ($K_{true}$) and applying Varimax rotation to maximize separability \citep{Kaiser1958}, thereby establishing a baseline for attainable accuracy under ideal - albeit unrealistic - supervised conditions.

\subsection{Performance metrics}
The quality of the detected partitions $\mathcal{C}_{det}$ against the ground truth $\mathcal{C}_{true}$ is evaluated using three complementary metrics, including classification accuracy, internal reliability, and topological quality. 

Classification Accuracy, measures the topological recovery using the Adjusted Rand Index (ARI) \citep{Steinley2004}. The ARI quantifies the similarity between two partitions, corrected for chance. An $\text{ARI} = 1$ indicates a perfect reconstruction of the latent themes, while $\text{ARI} \approx 0$ implies random assignment. On the other hand, internal reliability ($\alpha$) ensures that the topological clusters translate into valid psychometric instruments, we compute Cronbach's $\alpha$ for each detected community $c$. Given the rank-deficiency of the covariance matrix ($N < P$), we employ smoothed covariance estimation to compute:
    \begin{equation}
       \alpha_c = \frac{P_c}{P_c-1} \left( 1 - \frac{\sum_{j \in c} \sigma^2_j}{\sigma^2_{total}} \right)
    \end{equation}
\noindent where \(P_c\) is the number of items.
    
 Finally, the topological quality ($Q$), assesses the structural density of the partitions using the Modularity score $Q$ (independent of the ground truth). This metric serves as a proxy for the internal cohesion of the consensus, verifying whether the algorithm has successfully optimized the network topology. However, to evaluate the robustness of the algorithms beyond the baseline simulation, we implemented a sensitivity analysis protocol. We systematically increased the inter-cluster noise level $\rho_{out}$ from $0.05$ to $0.40$ (in steps of $0.05$) while holding the intra-cluster signal constant ($\rho_{in} = 0.50$). For each noise increment, we performed Monte Carlo iterations to compute the decay in classification accuracy. This "stress test" is designed to identify the breakdown point of each topological approach, distinguishing between algorithms that offer high resolution in ideal conditions versus those that maintain structural stability in high-entropy regimes.
Collectively, these metrics provide a multidimensional assessment, ensuring that the selected algorithm not only reconstructs the latent topology with high fidelity but also yields psychometrically sound and structurally robust consensus partitions.

\subsection{Ground truth recovery and classification accuracy}

In this section, we present the benchmarking results of the proposed topological clustering framework presented in Section \ref{sec:methodology}. Applying the simulation protocols previously defined, we evaluate the performance of the four algorithms - Louvain, Leiden, Infomap, and Spectral Clustering - across four dimensions: (1) Ground truth recovery accuracy; (2) Topological flow and stability; (3) Structural quality (Modularity); and (4) Internal reliability (Consistency).

The primary benchmark metric, the ARI, quantifies the similarity between the detected partitions and the latent ground truth, corrected for chance. Table \ref{tab:ari_results} summarizes the classification performance.

\begin{table}[h]
\centering
\caption{Comparative performance of community detection algorithms ($N=20, P=60$).}
\label{tab:ari_results}
\begin{tabular}{lccc}
\hline
\textbf{Algorithm} &  & \textbf{Detected $K$} & \textbf{ARI Score} \\ \hline
Louvain &  & 3 & 0.755 \\
Leiden &  & 3 & 0.794 \\
Spectral Clustering &  & 3 & 0.769 \\
Infomap &  & 4 & \textbf{0.839} \\ \hline
\end{tabular}
\end{table}

As shown in Table \ref{tab:ari_results}, all algorithms achieved robust reconstruction of the latent structure ($\text{ARI} > 0.75$), validating the efficacy of the network approach in small-sample regimes.
Infomap achieved the highest classification accuracy ($\text{ARI} = 0.839$). However, this superiority in item-level assignment came with a topological divergence: Infomap detected $K=4$ communities instead of the nominal $K_{true}=3$. This suggests that while Infomap is highly precise in classifying core items, its flow-based objective function (the Map Equation) is sensitive to substructures, leading to the segmentation of a single latent theme into two distinct communities.

\begin{figure}[H]
\centering
\setlength{\fboxsep}{0pt}
\setlength{\fboxrule}{0.3pt}
\fbox{\includegraphics[width=0.98\textwidth]{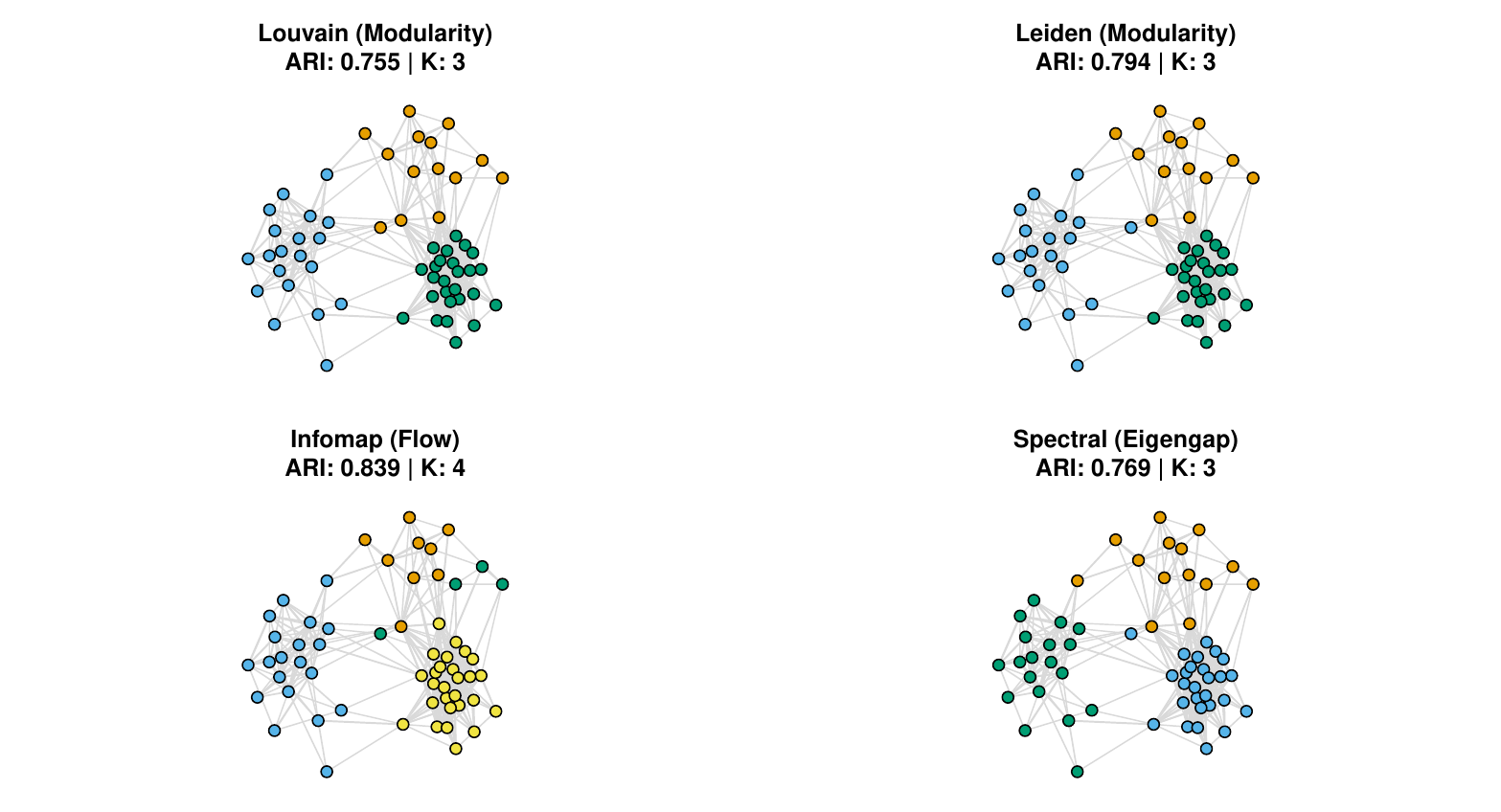}}
\caption{Comparative visualization of consensus networks. Nodes represent items, and edges represent significant correlations ($p < .05$, FDR corrected). The layout is fixed across all plots to facilitate comparison. Colors indicate community membership detected by: (1) Louvain, (2) Leiden, (3) Infomap, and (4) Spectral Clustering.}
\label{fig:network_comparison}
\end{figure}

To provide a qualitative assessment of the partition structures, we visualized the consensus networks generated by each algorithm (Figure \ref{fig:network_comparison}). The graphs were rendered using the Fruchterman-Reingold force-directed layout \citep{Gajdos2016}, which positions nodes based on their connectivity strength: highly correlated items are spatially proximal, while weakly associated items are pushed apart. The visual inspection corroborates the quantitative findings. Louvain, Leiden, and Spectral clustering (Figure \ref{fig:network_comparison}: plots 1, 2, and 4) exhibit a striking structural isomorphism. They all identify three distinct, dense macro-communities (represented by the blue, green, and orange/yellow regions), confirming that density-based and spectral methods converge on the same global topology.

In contrast, the Infomap topology (Figure \ref{fig:network_comparison}: plot 3) explicitly visualizes the phenomenon of over-segmentation. While the upper and left clusters remain identical to the other methods, the dense cluster in the bottom-right quadrant is partitioned into two distinct sub-communities (yellow and green). Crucially, this split is not spatially random; the two sub-groups appear to occupy distinct regions within the larger cluster, suggesting that Infomap is detecting a latent "bottleneck" in the information flow that separates two closely related but distinct sub-themes. This visual evidence supports the interpretation that the fourth cluster is a genuine topological feature rather than an artifact of noise.

Among the modularity-based approaches - as highlighted in Table \ref{tab:ari_results} - Leiden ($\text{ARI} = 0.794$) outperformed the classical Louvain algorithm ($\text{ARI} = 0.755$). Both correctly identified the number of communities ($K=3$). The superior performance of Leiden confirms its theoretical advantage in resolving the "badly connected communities" problem inherent in Louvain's heuristic, providing a more accurate recovery of the unbalanced clusters, particularly the smaller peripheral group ($n=10$). Moreover, spectral clustering, guided by the eigengap heuristic, also correctly identified $K=3$ with an accuracy ($\text{ARI} = 0.769$) comparable to the modularity methods, demonstrating the robustness of the spectral gap in noisy, ordinal environments.

\subsection{Analysis of algorithmic divergence and modularity assessment}

\begin{figure}[h]
\centering
\setlength{\fboxsep}{0pt}
\setlength{\fboxrule}{0.3pt}
\fbox{\includegraphics[width=0.98\textwidth]{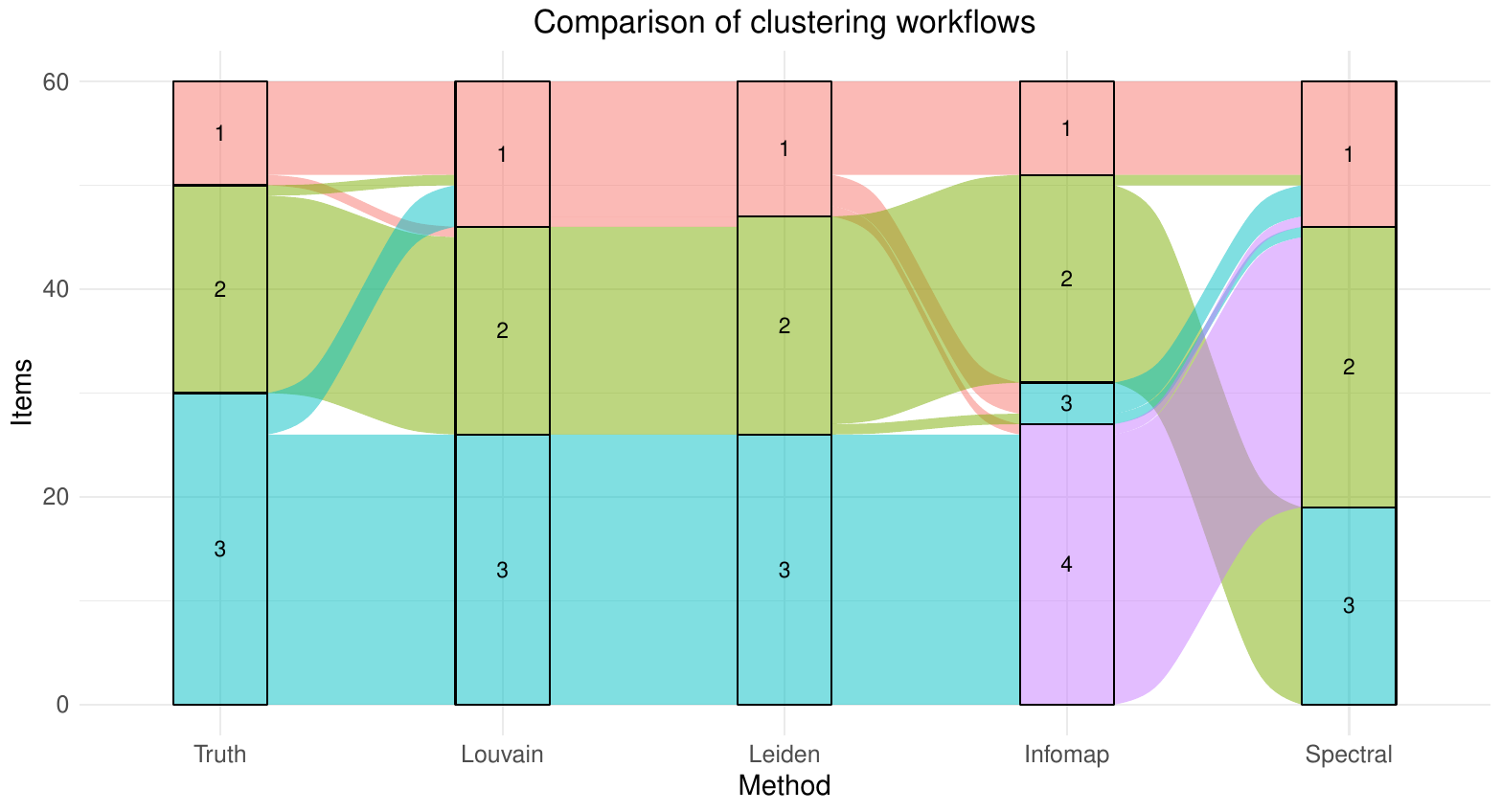}}
\caption{Alluvial diagram illustrating the flow of item classification across clustering workflows. Note the bifurcation of the central cluster in the Infomap column.}
\label{fig:alluvial}
\end{figure}

To investigate the structural reasons behind the divergence in ARI scores and the detected number of clusters ($K$), we employed an alluvial diagram (Figure \ref{fig:alluvial}) to track the flow of item assignments across the different methods.

The Alluvial diagram reveals a stable consensus core across Louvain, Leiden, and Spectral methods, represented by horizontal flows that largely preserve the three-block structure of the Ground Truth.
However, a critical topological shift occurs with Infomap. As visualized in the transition from Leiden to Infomap, the flow corresponding to the second latent cluster (green band) undergoes a bifurcation. Infomap isolates a subset of items into a distinct fourth community (cyan block, labeled: 3), while retaining the majority in the original cluster (green block, labeled: 2).

This "over-segmentation" is evident, and while Leiden aggregates the items into a cohesive block to maximize structural density, Infomap's random walker detects a bottleneck in the information flow within this cluster. Consequently, it partitions the group to minimize the description length of the trajectory. In the context of Delphi studies, this result implies that Infomap acts as a fine-grain detector, capable of revealing latent sub-themes or "minority reports" that density-based algorithms (Leiden) might amalgamate into broader, more generalized categories.

Beyond ground truth recovery, we assessed the internal structural quality of the partitions using the Modularity score ($Q$). This metric evaluates the strength of the division, independent of external validation.

\begin{table}[h]
\centering
\caption{Topological quality of partitions (Modularity $Q$).}
\label{tab:modularity}
\begin{tabular}{lc}
\hline
\textbf{Algorithm} & \textbf{Modularity ($Q$)} \\ \hline
\textbf{Leiden} & \textbf{0.4261} \\
Louvain & 0.4248 \\
Infomap & 0.4201 \\
Spectral Clustering & 0.4177 \\ \hline
\end{tabular}
\end{table}

The results in Table \ref{tab:modularity} confirm the theoretical expectations. Leiden achieved the highest modularity ($Q=0.4261$), marginally outperforming Louvain ($Q=0.4248$). This reinforces Leiden's status as the optimal choice when the analytical goal is to maximize the global cohesion and separability of the consensus themes.
Notably, Infomap yielded a lower modularity score ($Q=0.4201$). This reduction is mathematically consistent with the over-segmentation observed in the ARI analysis. By splitting a dense thematic cluster to optimize flow, Infomap inherently sacrifices a fraction of the global modularity. However, the minimal difference ($\Delta Q \approx 0.006$) suggests that the fourth community identified by Infomap is topologically plausible and not merely an artifact of noise.

\begin{figure}[h]
\centering
\setlength{\fboxsep}{0pt}
\setlength{\fboxrule}{0.3pt}
\fbox{\includegraphics[width=\textwidth]{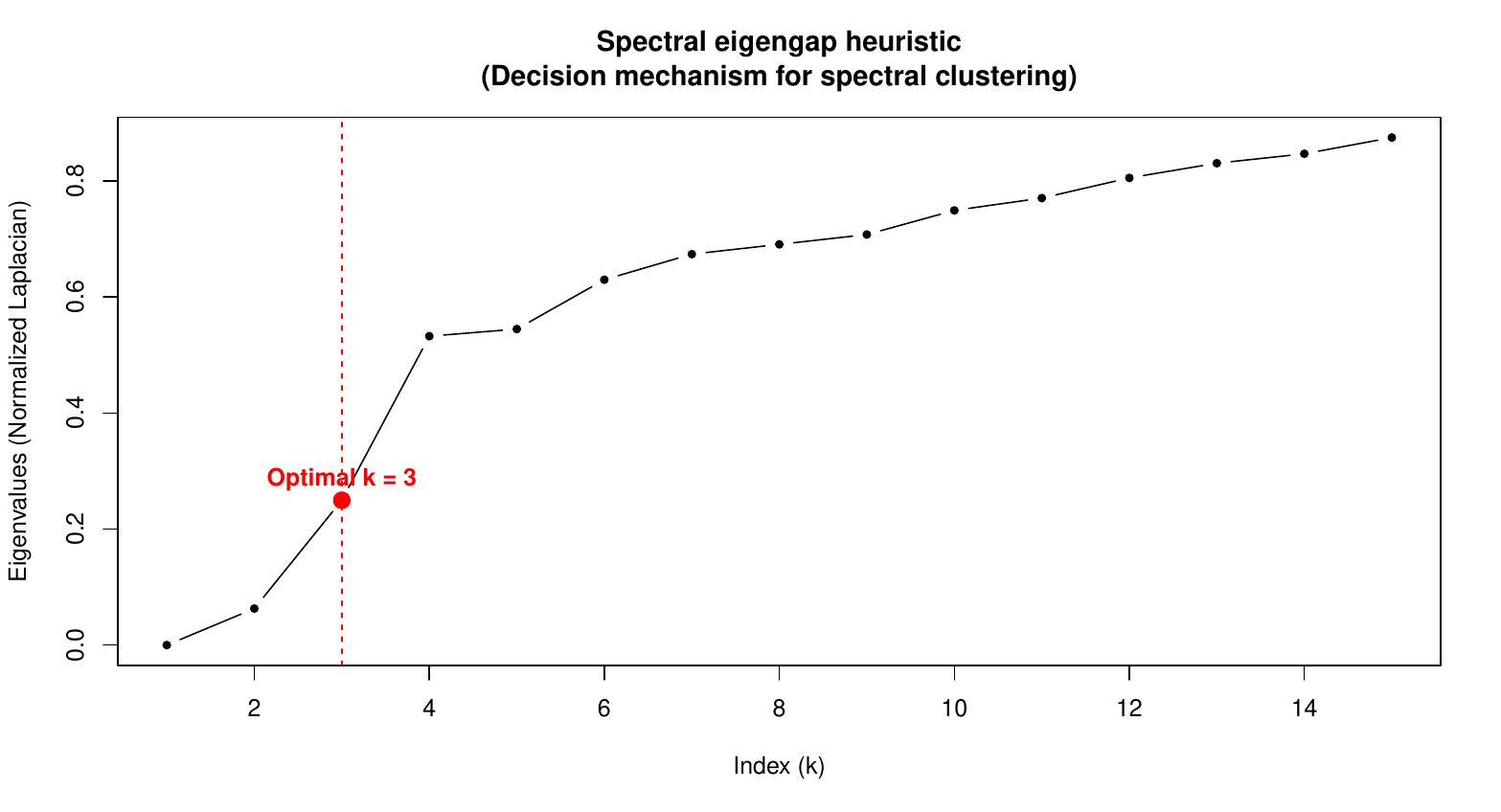}}
\caption{Spectral Eigengap Heuristic. The optimal number of clusters $K=3$ is identified by the maximum difference between consecutive eigenvalues ($\delta_k = |\lambda_{k+1} - \lambda_k|$).}
\label{fig:eigengap}
\end{figure}

Finally, the performance of the spectral clustering algorithm relies heavily on the correct estimation of $K$. Figure \ref{fig:eigengap} illustrates the Eigengap heuristic applied to the Normalized Laplacian of the consensus network.

The plot exhibits a distinct "elbow" or spectral gap at index $K=3$. The first three eigenvalues are near zero, indicating three connected components (or strongly coupled clusters) in the graph spectrum. The jump to the fourth eigenvalue ($\lambda_4$) represents the energy required to cut the graph further. The algorithm's ability to correctly identify $K=3$ despite the rank-deficiency of the input matrix ($N=20$) validates the use of spectral methods as a robust, unsupervised decision mechanism for determining the dimensionality of Delphi questionnaires.

\subsection{Internal consistency and robustness}
\begin{figure}[h]
\centering
\setlength{\fboxsep}{0pt}
\setlength{\fboxrule}{0.3pt}
\fbox{\includegraphics[width=\textwidth]{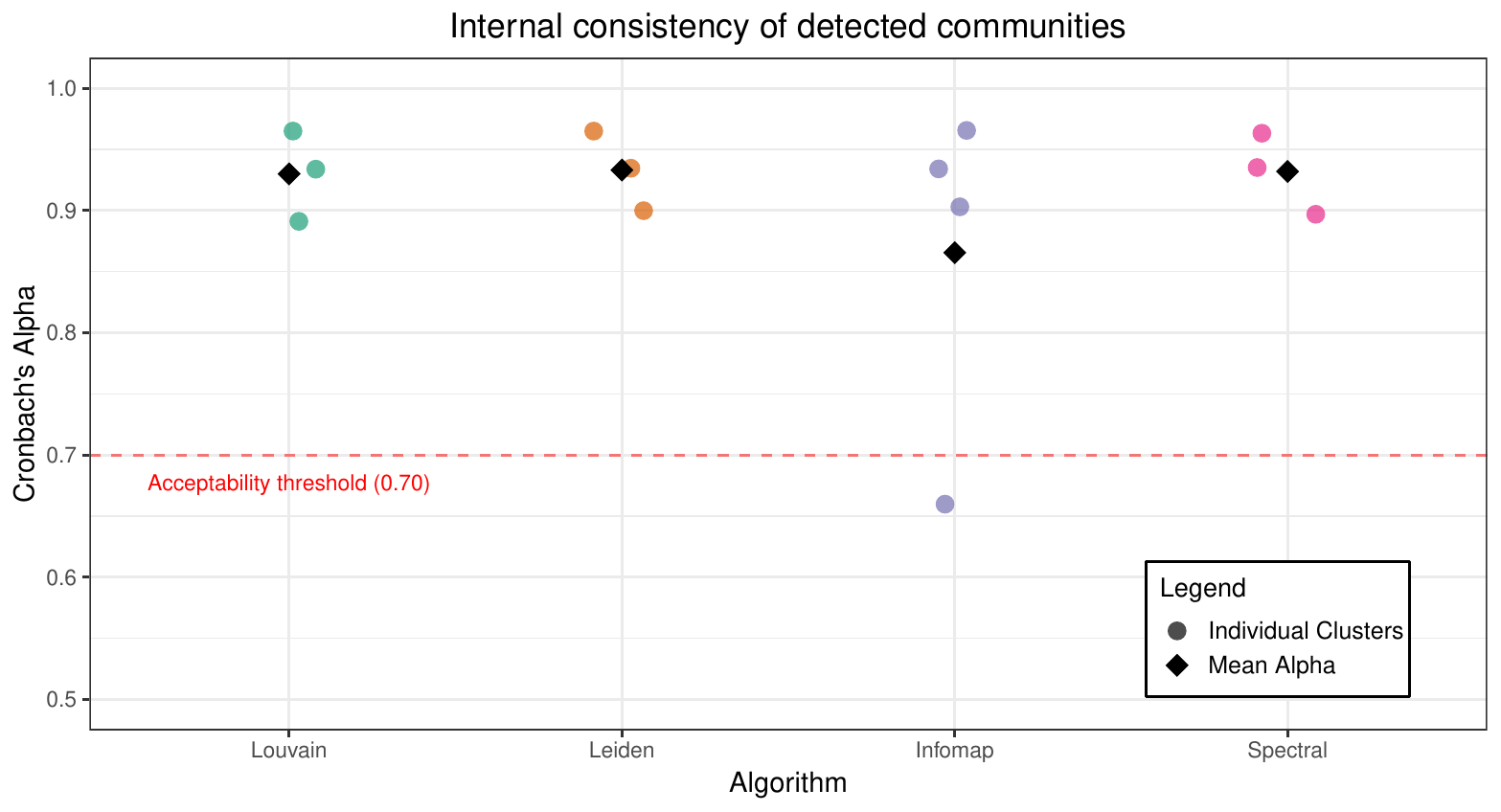}}
\caption{Strip plot of Internal Consistency (Cronbach's $\alpha$) for detected communities. Black diamonds represent the mean $\alpha$ per algorithm; colored dots represent individual clusters. The red dashed line indicates the acceptability threshold ($\alpha=0.70$).}
\label{fig:alpha}
\end{figure}

A critical requirement for Delphi analysis is that the identified clusters must correspond to valid measurement scales. We evaluated this using Cronbach's $\alpha$ for each detected community (Figure \ref{fig:alpha}). The analysis provides decisive evidence regarding the validity of the partitions:
\begin{itemize}
    \item \textit{High-reliability consensus}: firstly, the algorithms based on density (Louvain, Leiden) and spectral partitioning produced clusters with exceptional internal consistency. As shown in Figure \ref{fig:alpha}, all data points for these methods cluster tightly in the upper region of the plot, with $\alpha$ values ranging between $0.89$ and $0.96$ (Mean $\alpha \approx 0.93$). This confirms that the topological clusters represent coherent "belief systems" where experts show high inter-item agreement.
    \item \textit{The Infomap trade-off}: secondly, the distribution for Infomap reveals the consequence of its over-segmentation. While three of its clusters maintain high reliability ($\alpha > 0.90$), the fourth isolated cluster appears as an outlier with $\alpha \approx 0.66$.
\end{itemize}

This finding is significant, since an $\alpha$ of $0.66$, while below the strict $0.70$ threshold, indicates a "marginal but existing" consistency. It suggests that Infomap did not create a random cluster (which would have $\alpha \approx 0$), but rather isolated a sub-dimension with weaker internal coherence. This supports the interpretation of Infomap as a tool for detecting fringe consensus or niche disagreements, whereas Leiden is superior for establishing the core consensus.

\begin{figure}[h]
\centering
\setlength{\fboxsep}{0pt}
\setlength{\fboxrule}{0.3pt}
\fbox{\includegraphics[width=\textwidth]{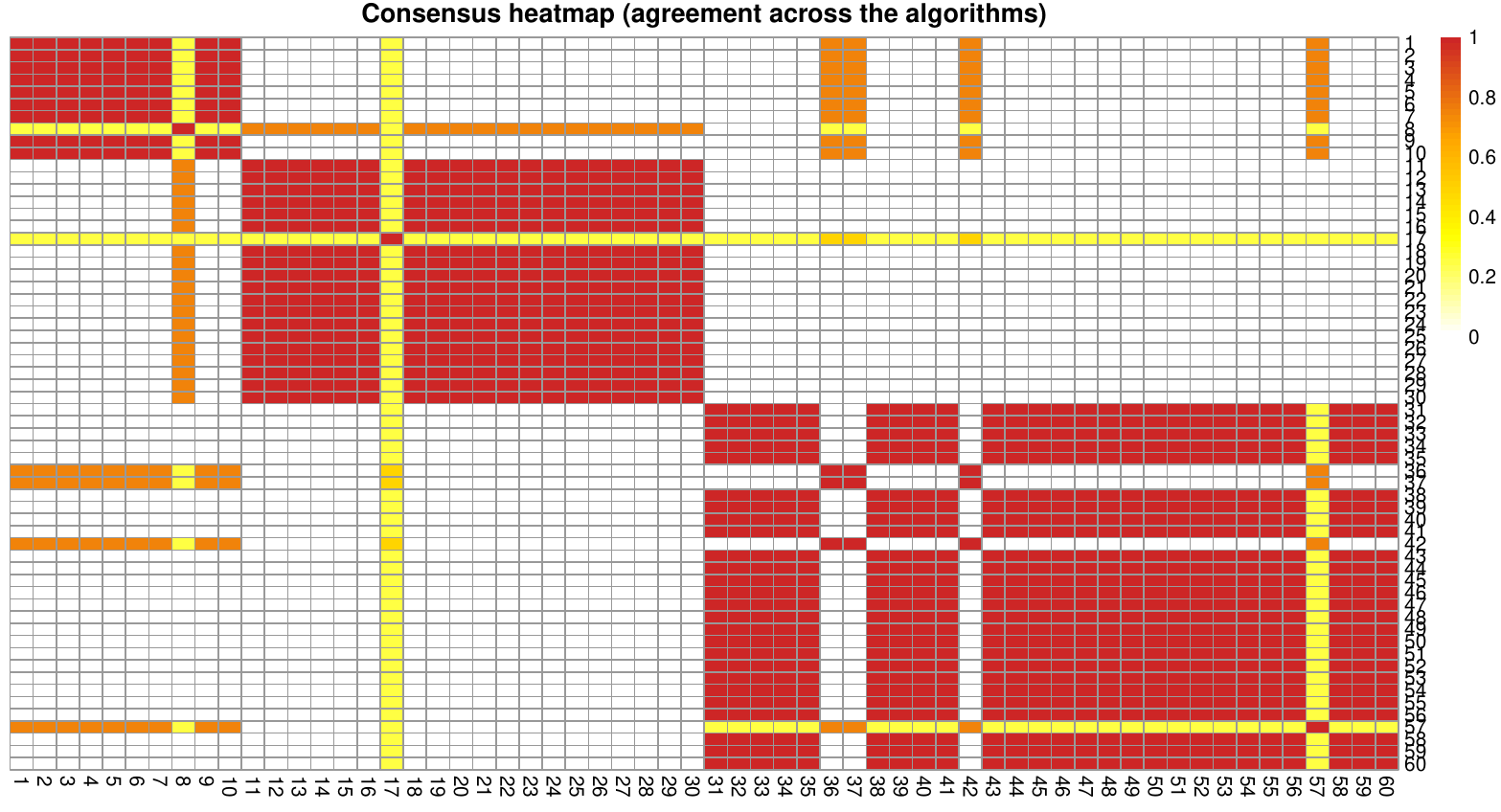}}
\caption{Consensus Heatmap aggregating partitions from all four algorithms. Dark red blocks indicate 100\% agreement (item pairs always clustered together); yellow/orange bands indicate areas of algorithmic disagreement.}
\label{fig:heatmap}
\end{figure}

\begin{figure}[H]
\centering
\setlength{\fboxsep}{0pt}
\setlength{\fboxrule}{0.3pt}
\fbox{\includegraphics[width=\textwidth]{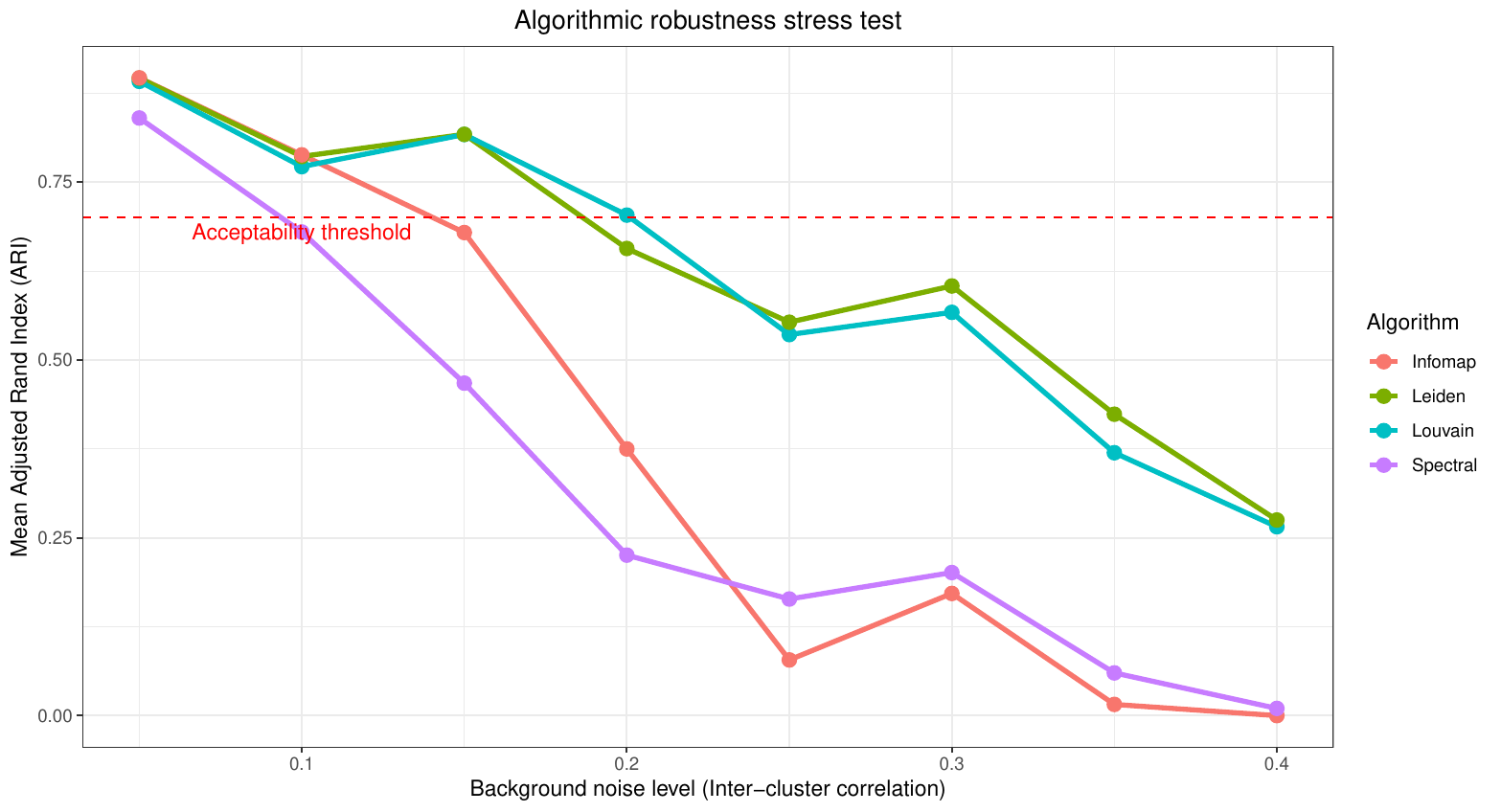}}
\caption{Algorithmic Stress Test (Sensitivity Analysis). The plot illustrates the decay in classification accuracy (Mean ARI) as a function of increasing systemic noise (inter-cluster correlation $\rho_{out}$). The dashed red line represents the reliability threshold ($\text{ARI}=0.70$). Note the divergent trajectories: Infomap shows a steep decay in high-noise regimes, while Leiden maintains robust stability.}
\label{fig:stress}
\end{figure}

To better investigate the algorithm consensus, we integrated the results of all four algorithms into a consensus heatmap (Figure \ref{fig:heatmap}) to visualize the stability of item associations. The heatmap demonstrates a high degree of topological stability. The diagonal is dominated by three solid, dark-red blocks, corresponding to the latent Ground Truth clusters. These regions represent the "core consensus", where all algorithms - regardless of their mathematical foundation (flow, density, spectrum) - agree on the item clustering.
The uncertainty is strictly confined to specific off-diagonal bands (visualized in orange/yellow). These "checkerboard" patterns correspond precisely to the items involved in the Infomap split (Cluster 2 vs. Cluster 3 in the Infomap partition). The absence of scattered noise outside these bands confirms that the divergence between methods is systematic and structurally driven, rather than random. This allows researchers to confidently accept the core themes while flagging the orange regions for qualitative review by the expert panel. To explicitly quantify the resilience of the topological models against systemic noise, we performed a stress test by varying the inter-cluster correlation ($\rho_{out}$) from a baseline of $0.05$ (near-perfect signal) to $0.40$ (high entropy), while keeping the intra-cluster signal fixed (also explained in Figure \ref{fig:stress})

Considering the stress test performed by the sensitivity analysis, the trajectories reveal a fundamental functional trade-off between "Resolution" and "Robustness" (Figure \ref{fig:stress}). In particular as hypothesized, Infomap (represented by the steepest curve) exhibits the behavior of a high-sensitivity instrument. In low-noise regimes ($\text{Noise} < 0.15$), it matches or exceeds the other algorithms, adopting the Map Equation to detect fine-grained substructures. However, its performance degrades rapidly as the signal-to-noise ratio decreases. The steep negative slope indicates that flow-based dynamics are highly susceptible to "pathway contamination": as noise introduces spurious edges between clusters, the random walker loses confinement, causing the partition quality to drop below the acceptability threshold ($\text{ARI} < 0.70$) significantly earlier than density-based methods.

Conversely, the modularity-based algorithms (particularly Leiden) display a significantly flatter decay curve. Leiden maintains robust classification accuracy ($\text{ARI} > 0.75$) even in high-noise environments where Infomap falters. This confirms that Modularity maximization acts as a low-pass filter: by focusing on the aggregate density of connections rather than the specific trajectory of a random walker, Leiden effectively ignores local noise perturbations, preserving the global macro-structure of the consensus.

Finally, spectral clustering algorithm mirrors the robustness profile of Leiden, further validating the use of the Eigengap heuristic. The stability of the eigenvalues against perturbation ensures that the dimensionality estimation ($K$) remains correct until the noise overwhelms the spectral gap. In conclusion, this stress test provides the empirical rationale for the proposed methodological framework: while Infomap offers superior granularity for clean, high-consensus datasets, Leiden provides the necessary safety margin for ambiguous or highly noisy Delphi panels.

\begin{figure}[H]
\centering
\setlength{\fboxsep}{0pt}
\setlength{\fboxrule}{0.3pt}
\fbox{\includegraphics[width=\textwidth]{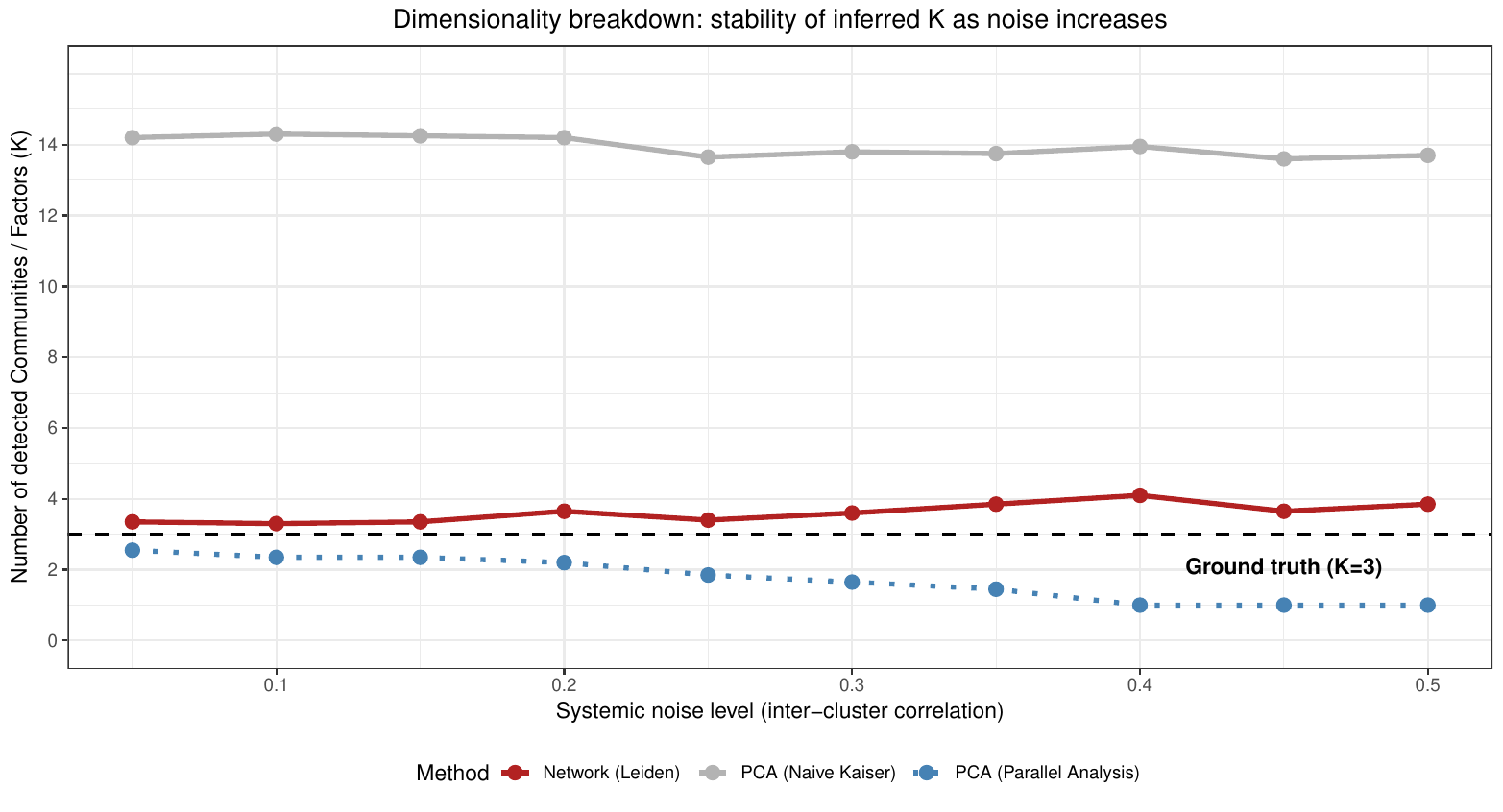}} 
\caption{Dimensionality breakdown analysis. The plot contrasts the stability of the inferred number of clusters ($K$) as systemic noise increases. While the Naive PCA (Kaiser) massively overestimates dimensionality and Parallel Analysis exhibits early sensitivity to noise, the Network approach (Leiden) maintains the correct ground truth ($K=3$) across a wider range of signal-to-noise ratios.}
\label{fig:stability_k}
\end{figure}

The comparative analysis reveals a critical divergence in structural recovery capabilities (Table \ref{tab:pca_benchmark}). The Kaiser-Meyer-Olkin test yielded a value of $\text{MSA} = 0.50$, placing the dataset on the threshold of inadmissibility for factor analysis. Crucially, the limitations of variance-based models became evident in the dimensionality assessment. While the network-based Leiden algorithm autonomously identified the correct dimensionality ($K=3$), the standard Kaiser criterion for PCA suggested the extraction of $K=14$ factors, conflating sampling noise with latent signal. Even when employing the more robust Parallel Analysis (PA), which correctly identified the dimensionality ($K=3$), the resulting factor structure remained statistically unstable due to the inadmissibility of the correlation matrix ($\text{MSA} < 0.50$). Although the simulation setting is challenging for variance-based methods, PCA is evaluated here under a favorable assumption in which the true number of clusters is fixed a priori ($K=K_{true}$). Yet, this represents a supervised scenario unavailable in practice. Despite this advantage, when the constraint is removed, PCA shows a degradation in performance. In contrast, the topological approach provided a robust, "turn-key" solution ($\text{ARI} \approx 0.794$) without requiring a priori knowledge of the latent structure, validating its superiority in blind small-sample scenarios.
\begin{table}[h]
\centering
\small 
\caption{Benchmark: Network Approach (Leiden) vs. Traditional PCA ($N=20, P=60$).}
\label{tab:pca_benchmark}
\begin{tabularx}{\textwidth}{ >{\raggedright\arraybackslash}p{3.8cm} X X }
\toprule
\textbf{Metric} & \textbf{Network Approach} & \textbf{PCA / Factor Analysis} \\
\midrule

Mathematical Suitability \newline (Metric Value) & 
Modularity ($Q$) \newline 0.426 (Good structure) & 
KMO Test (MSA) \newline 0.500 (Weak) \\ 
\addlinespace

Dimensionality Detection \newline (Criterion) & 
$K = 3$ (Correct) \newline Modularity Optimization & 
1) $K = 14$ (Naive Kaiser) \newline 2) $K = 3$ (Parallel Analysis)* \\ 
\addlinespace

Classification Accuracy \newline (Score) & 
Adjusted Rand Index \newline 0.794 (Unsupervised) & 
Adjusted Rand Index \newline 0.833 (Forced to $K_{true}$) \\ 
\addlinespace

Interpretation & 
Stable \& Robust & 
Matrix Ill-Conditioned \\ 

\bottomrule
\multicolumn{3}{l}{\footnotesize *Parallel Analysis identified correct K, but factors are unstable (KMO < 0.5).}
\end{tabularx}
\end{table}
Finally, to rigorously justify the proposed method, we evaluated the "breakdown point" of the dimensionality inference. Figure \ref{fig:stability_k} illustrates the stability of the detected number of clusters ($K$) as a function of increasing systemic noise. The results highlight a striking divergence in structural resilience.
Traditional PCA metrics exhibit dichotomous failure modes: the Naive Kaiser criterion systematically overestimates the structure (treating noise as factors, $K \approx 14$), whereas the robust Parallel Analysis suffers from conservative collapse. As shown by the dotted blue trajectory, PA tends to underestimate the latent complexity in high-noise regimes, effectively losing the signal of distinct themes as inter-cluster correlations rise.
In sharp contrast, the topological approach (Leiden, red trajectory) demonstrates superior structural fidelity. The network algorithm successfully anchors the consensus near the ground truth ($K \approx 3$) across the entire noise spectrum. This confirms that topological modularity acts as a stabilizing filter, capable of distinguishing latent themes even when the covariance signal becomes too ambiguous for factor analytic methods.

\subsection{Sensitivity to sample size and varying cluster dimensionality}
\label{subsec:sensitivity_nk}

To ensure our findings are not artifacts of a single favorable baseline, we expanded the simulation to investigate the framework's behavior across varying $N/P$ concentration ratios ($N \in \{15, 30, 60\}$ for $P=60$) and varying structural complexities ($K_{true} \in \{2, 5\}$). Crucially, to address potential alternatives to factor-analytic models, we introduced two distinct unsupervised benchmarks: (i) traditional Principal Component Analysis (PCA) operating under the standard Kaiser criterion (eigenvalues $> 1$) for blind dimensionality retention, and (ii) Hierarchical Agglomerative Clustering (HAC) using average linkage and a distance matrix computed as $1 - \rho_{\text{Spearman}}$, evaluated under an "oracle" setting forced to extract the exact true number of latent clusters ($K_{true}$).
The results, summarized in Table \ref{tab:sensitivity}, reveal a severe vulnerability in standard geometric and variance-based methods under small-sample constraints. In extreme sparsity regimes ($N/P = 0.25$), unsupervised PCA systematically collapses due to severe dimensionality overestimation, splitting cohesive groups into spurious components (Adjusted Rand Index as low as 0.241). Notably, even when hierarchical clustering algorithms are supplied with the true number of dimensions in advance, as in the case of HAC ($K_{true}$), traditional distance-based agglomeration struggles to resolve highly fragmented structures from noisy correlation matrices. For instance, at $N=15$ with $K=5$, HAC achieves an ARI of only 0.551.

\begin{table}[h]
\centering
\caption{Sensitivity analysis: Adjusted Rand Index (ARI) across varying sample sizes ($N$) and structural complexities ($K$) for $P=60$ items. PCA uses the Kaiser criterion; HAC uses average linkage and $1-\text{Spearman}$ distance at $K_{true}$.}
\label{tab:sensitivity}
\resizebox{\columnwidth}{!}{%
\begin{tabular}{ccccccc}
\toprule
\textbf{Structure} & \textbf{Experts ($N$)} & \textbf{$N/P$ Ratio} & \textbf{Leiden ARI} & \textbf{Spectral ARI} & \textbf{HAC ($K_{true}$) ARI} & \textbf{PCA ARI} \\
\midrule
$K=2$ & 15 & 0.25 & 0.455 & 0.702 & 0.683 & 0.241 \\
$K=2$ & 30 & 0.50 & 0.967 & 0.945 & 0.980 & 0.343 \\
$K=2$ & 60 & 1.00 & 0.980 & 0.991 & 1.000 & 0.560 \\
\midrule
$K=5$ & 15 & 0.25 & 0.613 & 0.466 & 0.551 & 0.461 \\
$K=5$ & 30 & 0.50 & 0.841 & 0.668 & 0.844 & 0.676 \\
$K=5$ & 60 & 1.00 & 0.944 & 0.960 & 0.996 & 0.846 \\
\bottomrule
\end{tabular}%
}
\end{table}

Conversely, the topological framework - specifically the Leiden algorithm - demonstrates remarkable resilience. By filtering out non-significant associations via the FDR thresholding mechanism prior to community extraction, the network protocol isolates the true structural modules without requiring manual or supervised calibration of $K$. In the most challenging sparse scenario ($N=15, K=5$), Leiden outperforms the oracle HAC benchmark, achieving an ARI of 0.613. As expected, as the sample size grows ($N=60$ and $N/P = 1.00$), the empirical correlation matrix stabilizes, and all non-parametric approaches converge toward excellent ground-truth recovery (ARI $\approx 0.99$). However, within the typical "Small $N$, Large $P$" domain of Delphi applications, the topological approach provides a distinctively robust, self-calibrating heuristic.

\section{Discussion and Conclusion}
\label{sec:conclusion}

This study addressed the longstanding statistical challenge of analyzing high-dimensional Delphi data under the constraints of small sample sizes ($N \ll P$). By shifting the analytical issue from latent variable models - which become statistically unstable due to rank-deficiency in small \(N\) large \(P\) settings - to a topological network framework, we demonstrated that community detection algorithms can effectively recover latent thematic structures from ordinal, noisy Delphi datasets.

The benchmarking analysis yielded robust evidence regarding the performance of topological clustering in rank-deficient regimes. First, all tested algorithms (Louvain, Leiden, Infomap, Spectral) achieved satisfactory ground-truth recovery ($\text{ARI} > 0.75$), validating the network approach as a reliable alternative to traditional factor analysis.
Specifically, Leiden emerged as the most structurally robust algorithm, achieving the highest modularity ($Q=0.426$) and correctly identifying the number of latent clusters ($K=3$). This confirms its suitability for maximizing the global density of consensus. Conversely, Infomap demonstrated superior item-level classification accuracy ($\text{ARI}=0.839$) but exhibited a tendency towards over-segmentation ($K=4$), isolating a peripheral sub-theme with lower psychometric reliability ($\alpha \approx 0.66$) compared to the core consensus blocks ($\alpha > 0.93$). Finally, the Spectral Eigengap heuristic proved effective in unsupervised dimensionality estimation, correctly identifying the structural break in the graph spectrum despite the limited sample size.

It is noteworthy that while the advanced Parallel Analysis correctly recovered the latent dimensionality ($K=3$), this result is deceptive in a small-sample context. With a KMO of $0.50$, the resulting factor solution lacks the psychometric stability required for reliable scale construction. In contrast, the network approach achieved the same dimensional accuracy but yielded topologically robust communities, proving that connectivity is a more reliable signal than covariance when $N \approx P$.

Our sensitivity analysis highlights a fundamental divergence between variance-based and topological approaches. Whereas PCA shows an inherent hypersensitivity to sampling noise, treating stochastic fluctuations as latent structures, community detection algorithms maintain structural resilience. Such stability represents a key methodological requirement for Delphi research, ensuring that thematic groupings reflect true consensus cohesion instead of small-sample mathematical artifacts.

From a theoretical perspective, this work establishes that collinearity in expert panels should not be treated as a statistical nuisance (singularity) to be regularized, but as a topological signal to be mapped. The primary methodological implication is that Network Analysis provides a mathematically viable approach for the "Small $N$, Large $P$" problem. By relying on pairwise topological similarities rather than requiring a well-conditioned global covariance structure, the graph-based approach allows researchers to retain the full complexity of the questionnaire without forcing variable reduction or violating asymptotic assumptions. Furthermore, the strong convergence between topological modularity and psychometric internal consistency (Mean $\alpha \approx 0.93$) suggests that community detection can serve as an automated, data-driven method for scale construction in exploratory research.

For practitioners and researchers employing the Delphi method, our findings suggest a tiered analytical protocol:
\begin{itemize}
    \item \textit{Core consensus definition}: the Leiden algorithm is recommended as the standard default. Its ability to guarantee connected communities and optimize structural density makes it ideal for defining the primary "chapters" or macro-themes of the expert consensus.
    \item \textit{Nuance and conflict detection}: Infomap should be employed as a complementary diagnostic tool. Its flow-based logic is uniquely sensitive to "minority reports" or subtle divergences in expert reasoning. If Infomap splits a cluster that Leiden keeps united, it signals a latent conceptual schism that warrants qualitative investigation, rather than statistical suppression.
    \item \textit{Dimensionality assessment}: the Spectral Eigengap offers a non-arbitrary, mathematical criterion to decide "how many themes exist", replacing subjective "rules of thumb" often used in qualitative analysis.
\end{itemize}

Furthermore, from an applicative point of view, the stability of the inferred dimensionality is vital for the interpretive utility of Delphi results. In the thematic clustering of questionnaire items, there is no fully objective criterion for determining the optimal number of groups; however, the literature suggests that generating an excessive number of clusters substantially reduces their analytical value \citep{Author2024}. While the set of possible thematic associations is theoretically vast, the number of clusters must remain deliberately small to preserve their function as a structured decision-support tool.
This principle is particularly evident in forecasting studies: for instance, when Delphi outcomes are used to build scenarios, producing more than three or four distinct visions often leads to cognitive overload for stakeholders, who may become "lost" in too many divergent perspectives. In this context, the tendency of traditional PCA to overestimate dimensionality (detecting up to 14 components in our simulation) represents a pragmatic failure. By maintaining a parsimonious and stable number of thematic clusters ($K \approx 3$), the proposed network framework ensures that the resulting synthesis remains actionable, focused, and cognitively manageable.
Moreover, once the optimal consensus communities are identified, the framework must bridge the gap between topological detection and practical interpretation. Depending on the research objective of the Delphi study, the features within each cluster can be synthesized using three distinct pathways: (1) Topological selection: rather than creating a synthetic variable, researchers can leverage network centrality metrics to extract the most representative item of a community. By calculating the within-module Node Strength (the sum of edge weights connecting a node to other nodes in the same cluster), the item with the highest strength can be objectively selected as the topological "centroid" or proxy for the entire theme. (2) Quantitative aggregation: as demonstrated by the high internal consistency (Cronbach's $\alpha$) of the detected modules, the communities function effectively as reliable psychometric scales. Consequently, the ordinal responses of the items within a cluster can be aggregated into a composite dimension (e.g., via the module median). This reduces the high-dimensional questionnaire into a few robust macro-indicators of consensus. (3) Qualitative abstraction: in foresight and policy-oriented Delphi studies, numerical aggregation is often secondary to narrative generation. The identified community serves as a semantic boundary: all items within the cluster are qualitatively synthesized by the research team into a unified "macro-scenario" or strategic narrative, ensuring that the resulting policy guidelines are grounded in statistically validated thematic cohesion.

\subsection{Limitations and future work}
Certain limitations of this study must be acknowledged. First and foremost, the validation was conducted exclusively on synthetic data. While the simulation was rigorously designed to mimic the pathological constraints of real consensus data (unbalanced clusters, ordinal discretization, systemic noise), the absence of a real-world empirical application limits the immediate demonstration of the framework in the field. Real-world human judgment may exhibit higher-order interactions, non-transitive preferences, or linguistic ambiguities not fully captured by the multivariate normal generation process. To illustrate how this framework bridges the gap between methodological benchmarking and practical decision-making, consider a prospective Delphi study tasked with forecasting the integration of Artificial Intelligence in the labor market. In such a scenario, experts might evaluate dozens of statements regarding automated decision-making, job displacement, and fuzzy boundaries between human and AI work. Using traditional variance-based methods, the resulting components would likely blend economic, technological, and ethical items into overlapping, continuous latent traits, forcing researchers to subjectively guess the cut-off points for action. By applying the proposed topological framework, the dense collinearity of expert judgments is natively partitioned into discrete, non-overlapping network communities (e.g., one strictly isolating "Ethical Guidelines," another grouping "Reskilling Policies"). This allows decision-makers to bypass statistical ambiguity and directly translate the mathematically objective clusters into targeted, coherent policy scenarios. Establishing this formal benchmark on synthetic data was a necessary first step to guarantee algorithmic stability; however, a critical avenue for future research is the deployment of this network protocol on empirical Delphi datasets to validate its scenario-building efficacy in vivo.

 Finally, this study focused on a static snapshot of the final consensus round. A promising avenue for future development is the application of Temporal Network Analysis to longitudinal Delphi data. By modeling the process as a multilayer temporal graph, researchers could track the trajectory of consensus formation, mathematically quantifying how expert opinions converge (or diverge) across iterative rounds.

\section*{Data availability}
The data that support the findings of this study are available from the corresponding author upon reasonable request.

\section*{Declaration of Competing Interest}
The authors declare that they have no known competing financial interests or personal relationships that could have appeared to influence the work reported in this paper.

\section*{Funding}
This research did not receive any specific grant from funding agencies in the public, commercial, or not-for-profit sectors.

\section*{CRediT authorship contribution statement}

\textbf{Author 1:} Conceptualization, Methodology, Software, Validation, Formal analysis, Investigation, Data Curation, Writing - Original Draft, Writing - Review \& Editing, Visualization. 
\textbf{Author 2:} Conceptualization, Methodology, Validation, Writing - Review \& Editing, Supervision.
\textbf{Author 3:} Conceptualization, Methodology, Validation, Writing - Review \& Editing, Supervision.

\bibliography{biblio}{}
\bibliographystyle{plainnat}

\end{document}